\documentclass[aps,prl,twocolumn,amsmath,amssymbs,superscriptaddress]{revtex4-1}

\usepackage{graphicx}
\usepackage{amsmath,amsfonts,amssymb}
\usepackage{epsfig}
\usepackage{wrapfig}
\usepackage{bbm}
\usepackage[usenames]{color}
\usepackage{array}
\usepackage{times}
\usepackage{setspace}
\usepackage{epstopdf}

\def\w{\omega}
\def\wo{\omega_{0}}
\def\wop{\omega_{0p}}
\def\wos{\omega_{0s}}
\def\woi{\omega_{0i}}
\def\wp{\omega_{p}}
\def\ws{\omega_{s}}
\def\wi{\omega_{i}}
\def\kin{\kappa_{in}}
\def\kex{\kappa_{ex}}
\def\a{a}
\def\ad{a^{\dag}}

\begin{document}

\title{A topological source of quantum light}
\author{Sunil Mittal}
\affiliation{Joint Quantum Institute, NIST/University of Maryland, College Park, MD 20742, USA}
\affiliation{Department of Electrical and Computer Engineering, and IREAP, University of Maryland, College Park, MD 20742, USA}
\author{Elizabeth A. Goldschmidt}
\affiliation{U.S. Army Research Laboratory, Adelphi, MD 20783, USA}
\affiliation{Joint Quantum Institute, NIST/University of Maryland, College Park, MD 20742, USA}
\author{Mohammad Hafezi}
\affiliation{Joint Quantum Institute, NIST/University of Maryland, College Park, MD 20742, USA}
\affiliation{Department of Electrical and Computer Engineering, and IREAP, University of Maryland, College Park, MD 20742, USA}
\affiliation{Department of Physics, University of Maryland, College Park, MD 20742, USA}


\begin{abstract}

Quantum light sources are characterized by their distinctive statistical distribution of photons. For example, single photons and correlated photon pairs exhibit antibunching and reduced variance in the number distribution that is impossible with classical light \cite{Shields2007, Eisaman2011}. Most common realizations of quantum light sources have relied on spontaneous parametric processes such as down-conversion (SPDC) and four-wave mixing (SFWM) \cite{Eisaman2011}. These processes are mediated by vacuum fluctuations of the electromagnetic field. Therefore, by manipulating the electromagnetic mode structure, for example, using nanophotonic systems, one can engineer the spectrum of generated photons \cite{Sharping2006, Clemmen2009, Fortsch2013, Davanco2012, Kumar2014}. However, such manipulations are susceptible to fabrication disorders which are ubiquitous in nanophotonic systems and lead to device-to-device variations in the spectrum of generated photons \cite{Topolancik2007, Mookherjea2008, Sapienza2010, Spring2017}. Here, we demonstrate topologically robust mode engineering of the electromagnetic vacuum fluctuations and implement a nanophotonic quantum light source where the spectrum of generated photons is robust against fabrication disorders. Specifically, we use the topological edge states to achieve an enhanced and robust generation of correlated photon pairs using SFWM and show that they outperform their topologically-trivial counterparts. We demonstrate the non-classical nature of our source using conditional antibunching of photons which confirms that we have realized a robust source of heralded single photons. Such topological effects, which are unique to bosonic systems, could pave the way for the development of robust quantum photonic devices.

\end{abstract}

\maketitle

SFWM is a third-order nonlinear process in which two pump photons at frequency $\wp$ are annihilated and two daughter photons, called signal and idler, are generated at frequencies $\ws$ and $\wi$. The spectrum of the generated signal and idler photons as well as their correlations are dictated by energy and momentum conservation relations, i.e., $2\wp = \ws + \wi$ and $2k_{p} = k_{s} + k_{i}$, where $k$'s  are the momenta of the respective fields. The spectrum is further constrained by the electromagnetic mode structure, i.e., the density of states or, equivalently, the dispersion relation $\w\left(k\right)$, which governs the propagation of the pump, signal and idler photons. Recently, the usage of nanophotonic systems, such as toroidal and ring resonators, has provided a compact and scalable route to manipulate the electromagnetic mode structure and hence, to implement spectrally engineered sources of correlated photons \cite{Eisaman2011, Sharping2006, Clemmen2009}. For example, coupled ring resonator arrays can be used to control the number of spectral modes \cite{Kumar2014} as well as to enhance the rate of photon pair generation, without compromising their bandwidth \cite{Davanco2012, Morichetti2011}. However, nanophotonic systems are invariably disposed to fabrication disorder which can significantly alter the dispersion of the photonic modes in an unpredictable fashion \cite{Topolancik2007, Mookherjea2008, Sapienza2010} and can result in randomness in the spectrum of photons generated by different devices. This randomness ultimately limits the scalability of such sources for practical applications in quantum communication and information processing which often require multiple sources with identical spectra, for example, in multi-photon interference scenarios \cite{Spring2017}.

At the same time, the introduction of topological protection in photonic systems has led to the development of a new class of devices which are inherently robust against disorder \cite{Ling2014, Wang2009, Hafezi2011, Hafezi2013, Rechtsman2013,  Chen2014, Cheng2016}. This robustness can be attributed to the presence of unidirectional, back-reflection free edge states in these systems. Edge states are characterized by topologically invariant integers \cite{Kraus2012, Hafezi2014, Mittal2016}, and therefore, photonic transport through these states is protected against local disorder \cite{Mittal2014, Wang2009, Cheng2016}. Edge states have been used to demonstrate, for example, robust optical delay lines \cite{Hafezi2013, Mittal2014}, reconfigurable photonic pathways \cite{Cheng2016}, topological lasers \cite{Jean2017, Bahari2017}. However, demonstrations of such topologically robust photonic systems have so far been confined to the classical regime.

In this work, we use topology for spectral engineering of the quantum fluctuations of the electromagnetic vacuum and implement a robust source of correlated photon pairs generated via SFWM. In particular, we exploit the linear dispersion associated with edge states for efficient phase-matching and show that the photon pair generation is significantly enhanced when the pump, as well as the signal and idler fields, correspond to edge modes of the system. We demonstrate correlations between the signal and idler photons beyond what is possible with classical sources and show conditional antibunching of photons, confirming the quantum nature of our source and its operation as a source of heralded single-photons. More importantly, using measurements over many devices, we show that the robustness of such topological spectral engineering manifests as a robustness in the spectrum of generated photons and our topological source outperforms a similarly-designed topologically-trivial source of correlated photons. From a fundamental perspective, our scheme is similar to theoretical proposals by Peano et. al. \cite{Peano2016} and Shi et. al. \cite{Shi2017} which investigated second- and third-order nonlinearity in topological edge states, respectively. These particle-nonconserving topological photonic systems have no counterparts in the electronic topological systems.

\begin{figure*}
\centering
\includegraphics[width=0.98\textwidth]{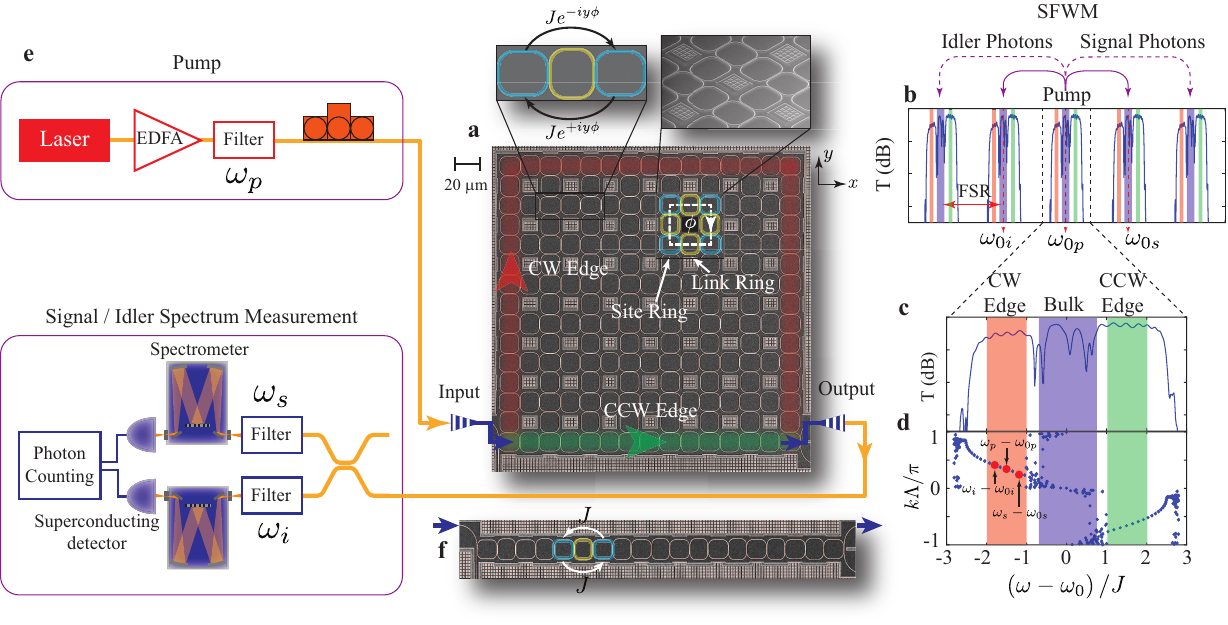}
\caption{
\textbf{Schematic of the experimental setup.} \textbf{a} SEM image of a 8$\times$8 lattice of site-ring resonators (cyan), coupled using link rings (yellow). Photons acquire a non-zero phase $\phi$ when then circulate around a plaquette of four site rings and four link rings. Insets show two site rings coupled by a link ring and a plaquette. The paths followed by clockwise (CW) and counter-clockwise (CCW) edge modes are highlighted in red and green, respectively. \textbf{b} The transmission spectrum of the device repeats after a free-spectral range (FSR). Correlated signal and idler photons are generated in longitudinal modes (of individual resonators) located symmetrically around the pump mode (centered at $\wop$). We choose the two modes one FSR above and below the pump mode, centered at frequencies $\wos$ and $\woi$, for collection of signal and idler photons. \textbf{c} Simulated transmission (T) spectrum of a 8$\times$8 lattice, in a given band. Two edge bands (shaded red and green) are separated by a bulk band (shaded blue). \textbf{d} Simulated dispersion curve showing linear dispersion for the edge modes. Here $k$ is the photon momentum and $\Lambda$ is the lattice constant such that $k\Lambda$ is the phase between two neighboring rings on the edge. Efficient phase-matching occurs when the pump as well as signal and idler frequencies correspond to edge modes. \textbf{e} Schematic of the pump and the spectral measurement setup. EDFA: erbium-doped fiber amplifier, PC: polarization controller. \textbf{f} SEM image of a topologically trivial 1D array of 10 site-ring resonators (cyan), coupled using link rings (yellow).
}
\end{figure*}

Our system consists of a 2D square lattice of ring resonators, positioned at the lattice sites, where the free non-interacting part of the photon dynamics is governed by the integer quantum-Hall model (Fig.1a) \cite{Hafezi2011, Hafezi2013}. A uniform synthetic magnetic field is synthesized by using link rings to couple to the neighboring site rings such that a photon hopping from one lattice site to its neighbor experiences a position- and direction-dependent hopping phase. The tight-binding Hamiltonian describing the linear evolution of the pump, signal and idler photons in the system is given as
\begin{eqnarray}
\nonumber H_{\text{L}} &=& \sum_{m,n} \omega_{0\mu} ~\ad_{m,\mu} \a_{m,\mu} \\
                       &-& J_{m,n} ~(\ad_{m,\mu} \a_{n,\mu} e^{-i \phi_{m,n}} + \ad_{n,\mu} \a_{m,\mu} e^{+i \phi_{m,n}}).
\end{eqnarray}
Here $\mu = p, s, i$ refers to the pump, signal or idler fields and $\ad_{m,\mu}$ is the corresponding photon creation operator at a lattice site $m = \left(m_{x},m_{y}\right)$, with frequency $\omega_{\mu}$. $J_{m,n}$ is the hopping rate of photons between lattice sites $m$, $n$ and is non-zero only for the nearest neighbor sites. $\phi_{m,n} = \phi ~m_{y} ~\delta_{m_{x},n_{x}+1} \delta_{m_{y},n_{y}}$ is the hopping phase between lattice sites and results in a uniform synthetic magnetic field flux $\phi$ per plaquette (Fig.1a). The energy spectrum of this Hamiltonian can be probed using transmission spectroscopy. For the chosen magnetic field strength $\phi = \frac{\pi}{2}$, the transmission spectrum consists of two edge bands at $\omega_{\mu} - \omega_{0\mu} \simeq \pm 1.5 J$, separated by a bulk band centered at $\omega_{\mu} - \omega_{0\mu} \simeq 0$ (Fig.1c). The edge bands are occupied by the topological edge states which are confined to the lattice boundary and circulate around the lattice in clockwise (CW) and counter-clockwise (CCW) directions, respectively (Fig.1a) \cite{Hafezi2013}. Furthermore, edge states are well described by a linear dispersion relation (Fig.1d) \cite{Hafezi2011, Mittal2016}. On the contrary, states in the bulk band occupy the bulk of the system and do not have a well-defined momentum. Note that this edge/bulk band structure repeats after every free-spectral-range (FSR), i.e., the frequency spacing between consecutive longitudinal modes of the individual ring resonators (Fig.1b). Consequently, the pump, signal and idler fields can occupy different longitudinal modes with resonance frequencies denoted by $\omega_{0\mu}$.

To generate correlated photon pairs in this system, we use the third-order nonlinearity of silicon and realize SFWM. This nonlinear four-photon interaction is described by the Hamiltonian
\begin{equation}
H_{\text{NL}} =  ~\eta \sum_{m}  \left( \ad_{m,s} ~ \ad_{m,i} ~ \a_{m,p} ~\a_{m,p} - \ad_{m,p} ~\ad_{m,p} ~\a_{m,s} ~\a_{m,i} \right),
\end{equation}
where $\eta$ is the strength of the SFWM and depends on the material and ring waveguide properties \cite{Chen2011, Ong2013}. The signal and idler modes are initially in the vacuum state when the input pump photons enter the system. However, the nonlinear interaction coherently adds or removes photon pairs from these vacuum modes and leads to generation of non-classical fields with intensity and spectral correlations between signal and idler photons \cite{Peano2016}. Furthermore, because of energy conservation, correlated signal and idler photon pairs are generated in longitudinal modes (of individual resonators) located symmetrically on either side of the pump mode \cite{Chen2011, Ong2013}. We choose signal and idler modes a single FSR above and below the pump mode with resonance frequencies denoted by $\wos, \woi ~\text{and} ~\wop$, respectively (Fig.1b). This choice allows us to effectively filter out the pump photons at the device output and also minimize the phase walk-off effects arising from the waveguide and material dispersion.

\begin{figure*}
\centering
\includegraphics[width=0.98\textwidth]{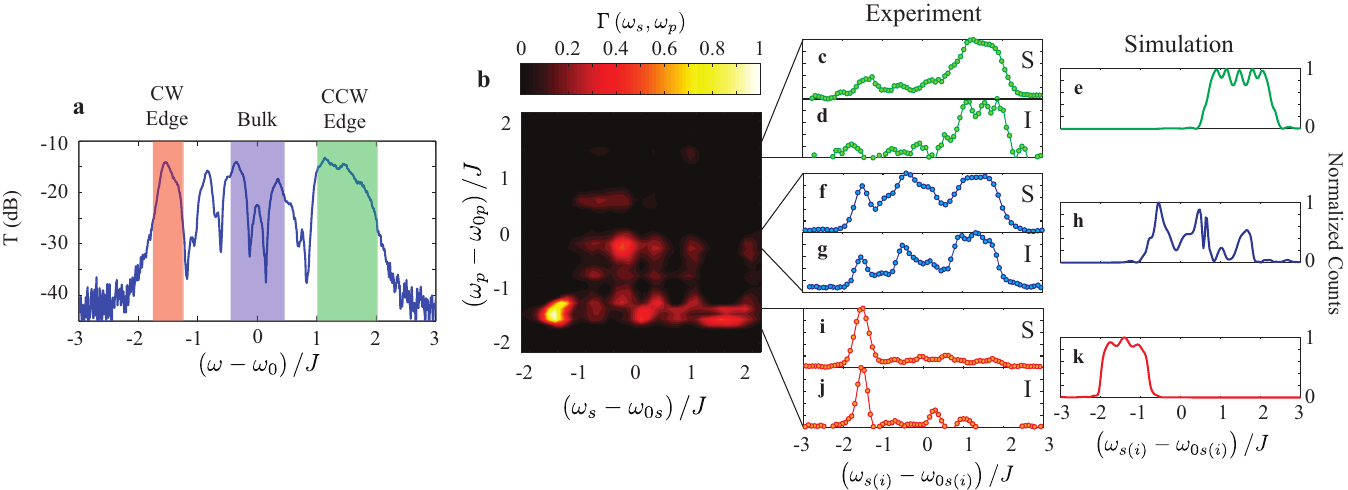}
\caption{
\textbf{Spectral distribution of the generated photons.} \textbf{a} Measured transmission spectrum for a $8 \times 8$ lattice. The edge and bulk bands are highlighted in color. \textbf{b} $\Gamma\left(\ws, \wp \right)$ - the intensity (normalized to unity) of generated signal photons at frequency $\ws$ as a function of pump frequency $\wp$. Maximum number of photons are generated when the pump as well as the signal and idler frequencies are in the CW edge band $\left(\w-\wo \simeq -1.5J \right)$. \textbf{c} Spectrum (normalized to unity) of signal (S) photons, i.e., horizontal cross-section of $\Gamma\left(\ws, \wp \right)$, at $\wp \simeq 1.5J$. With pump in CCW edge band, spectrum of generated signal photons is also limited to the CCW edge band. \textbf{d} Spectrum of idler (I) photons, with CCW edge band pump. Because of energy conservation, idler photons are also generated predominantly in the CCW edge band. \textbf{e} Simulation results for the spectrum of generated photons matches very well with experimental observations. \textbf{f-h} Corresponding results for pump in the bulk band. The signal and idler photons are generated throughout the spectrum of the lattice. Also, the simulation results do not match the observation because of the fabrication disorder in experimental system. \textbf{i-k} Signal and idler spectra when the system is pumped along the CW edge band, again showing spectrally confined generation of photons in the edge band.
}
\end{figure*}

In our experiment, we pump the lattice in one of the longitudinal modes using a tunable continuous-wave laser and measure the spectrum of generated photons. Figure 2a plots the linear pump transmission spectrum (see Extended Data Fig.E3 for more details) and Fig.2b plots $\Gamma\left(\ws, \wp \right)$ - the intensity of generated signal photons at frequency $\ws$ as we tune the pump frequency $\wp$. For a continuous-wave pump, measurement of $\Gamma\left(\ws, \wp \right)$ is equivalent to a measurement of the joint-spectral intensity which is commonly used to characterize the spectral correlations between generated photons (see Extended Data Fig.E4 and refs. \cite{Spring2017, Ong2013}). Firstly, we observe that the maximum number of photons are generated when the lattice is pumped in the CW edge band, at $\wp - \wop \simeq -1.5 J$. Secondly, with CW edge band pump, the spectrum of generated signal photons is predominantly confined to the CW edge band. This limited spectral distribution of signal photons can be seen more clearly with a normalized spectrum, the horizontal cross-section of $\Gamma\left(\ws, \wp \right)$, at $\wp - \wop \simeq -1.5J$ (Fig.2i). Furthermore, as a consequence of energy conservation, idler photons also exhibit a similar narrow spectrum centered at the CW edge band, i.e., $\wi - \woi = 2\left(\wp-\wop \right) - \left(\ws-\wos \right) \simeq -1.5 J$ (Fig.2j). This enhanced and spectrally limited generation of correlated photon pairs in the edge band is a result of the linear dispersion of edge modes which naturally satisfies the phase matching criteria and a good spatial overlap between the pump, signal and idler photons when they are confined to the lattice boundary. Our simulation results for the generated photon spectra agree well with our experimental observation (Fig.2k). We observe a similar, spectrally limited generation of correlated photons when the pump frequency is in the CCW edge band (Fig.2c-e). However, the propagation distance from the input to the output port is much shorter for the CCW edge modes compared to that of the CW edge modes and therefore, the intensity of generated photons is much weaker (Fig.1a).

In contrast to edge modes, bulk modes do not have a well-behaved dispersion (see Fig.1d) and their intensity distribution in the lattice changes even for very small changes in the excitation frequency \cite{Hafezi2013}. Therefore, in the bulk band, there is a significant phase mismatch between the pump, signal and idler photons, and their spatial overlap is also limited. As a result, the SFWM efficiency is low and photon pairs are generated throughout the transmission band of the lattice (Fig.2f-h). Moreover, the experimental and simulation results for the bulk band pump do not match. This is because our experimental system has fabrication disorder and the bulk band is not robust against disorder. On the contrary, a good agreement between the observed and simulated results for the edge states indicates their robustness against disorder.

\begin{figure}
\centering
\includegraphics[width=0.49\textwidth]{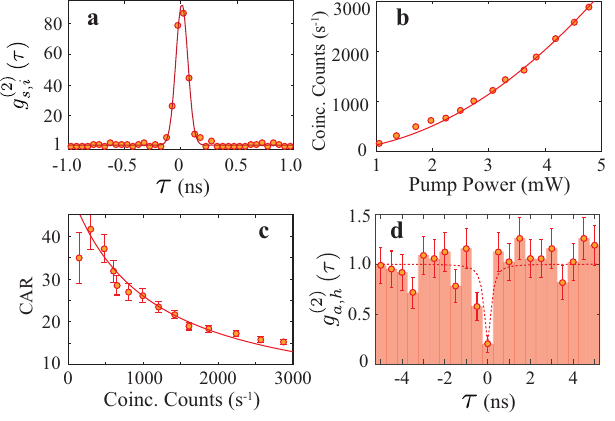}
\caption{
\textbf{Source characterization.} \textbf{a} Histogram for the cross-correlation function $g^{(2)}_{s,i} \left(\tau\right)$ between signal and idler photons, with a pump power of $\sim 2$ mW. \textbf{b} Coincidence count rate at the device output (adjusted for coupling losses), as a function of pump power. \textbf{c} CAR as a function of the coincidence count rate. \textbf{d} Histogram for the conditional (heralded) auto-correlation function $g^{(2)}_{a,h}$, with $g^{(2)}_{a,h}(0) = 0.20(8)$. The error bars in \textbf{b} and \textbf{c} are statistical. Solid lines are fit to the data and the dashed line is to guide the eye.
}
\end{figure}

To characterize the non-trivial nature of correlations between generated photons, we measure the second-order cross-correlation function, $g^{(2)}_{s,i} \left(\tau\right)$ which is the normalized probability of detecting signal and idler photons separated by time $\tau$ (see Methods and refs. \cite{Eisaman2011, Fortsch2013}). For two uncorrelated sources,  $g^{(2)} = 1$ for all $\tau$. In contrast, we observe a maximum $g^{(2)}_{s,i} \approx 80$ at $\tau = 0$ (Fig.3a). We integrate $g^{(2)}_{s,i} \left(\tau\right)$ over the peak at $\tau = 0$ to obtain the ratio of coincidence to accidental counts (CAR), which is analogous to the signal to noise ratio of a source. Our source achieves a CAR $\approx 42$ (Fig.3c) which is higher compared to similar other sources using single resonators \cite{Clemmen2009, Fortsch2013} and coupled resonators \cite{Davanco2012}, where CAR values of $\approx 30$ and $\approx 10$ were reported, respectively. This clearly indicates that the signal and idler photons are strongly correlated, i.e., the detection of a signal photon heralds the arrival of idler photon and vice-versa. Furthermore, we verified that the coincidence count rate between signal and idler photons increases as square of the pump power (Fig.3b) and CAR drops inversely with the coincidence rate (Fig.3c), as expected for SFWM interaction \cite{Sharping2006, Clemmen2009, Fortsch2013}.

Next, using a Hanbury Brown-Twiss setup, we measure the conditional auto-correlation function $g^{\left(2 \right)}_{a,h} \left(\tau \right)$ for signal photons, conditioned on the detection of idler photons (see Methods and refs. \cite{Eisaman2011, Fortsch2013}). Classical light sources are characterized by $g^{\left(2 \right)}_{a}\left(0 \right) \geq 1$ where the inequality holds for sources with bunched photons (such as thermal light), and $g^{\left(2 \right)}_{a}\left(0 \right) = 1$ when there are no correlations between arrival times of photons (as in lasers). Quantum light sources, such as single photons, are distinguished by $g^{\left(2 \right)} _{a}\left(0 \right) < 1$ which means that the photons are antibunched. We observe a conditional $g^{\left(2 \right)}_{a,h} \left(0 \right) = 0.20 \left(8 \right)$ which clearly shows antibunching and confirms that we have realized a topological source of heralded single photons (Fig.3d).

\begin{figure*}
\centering
\includegraphics[width=0.98\textwidth]{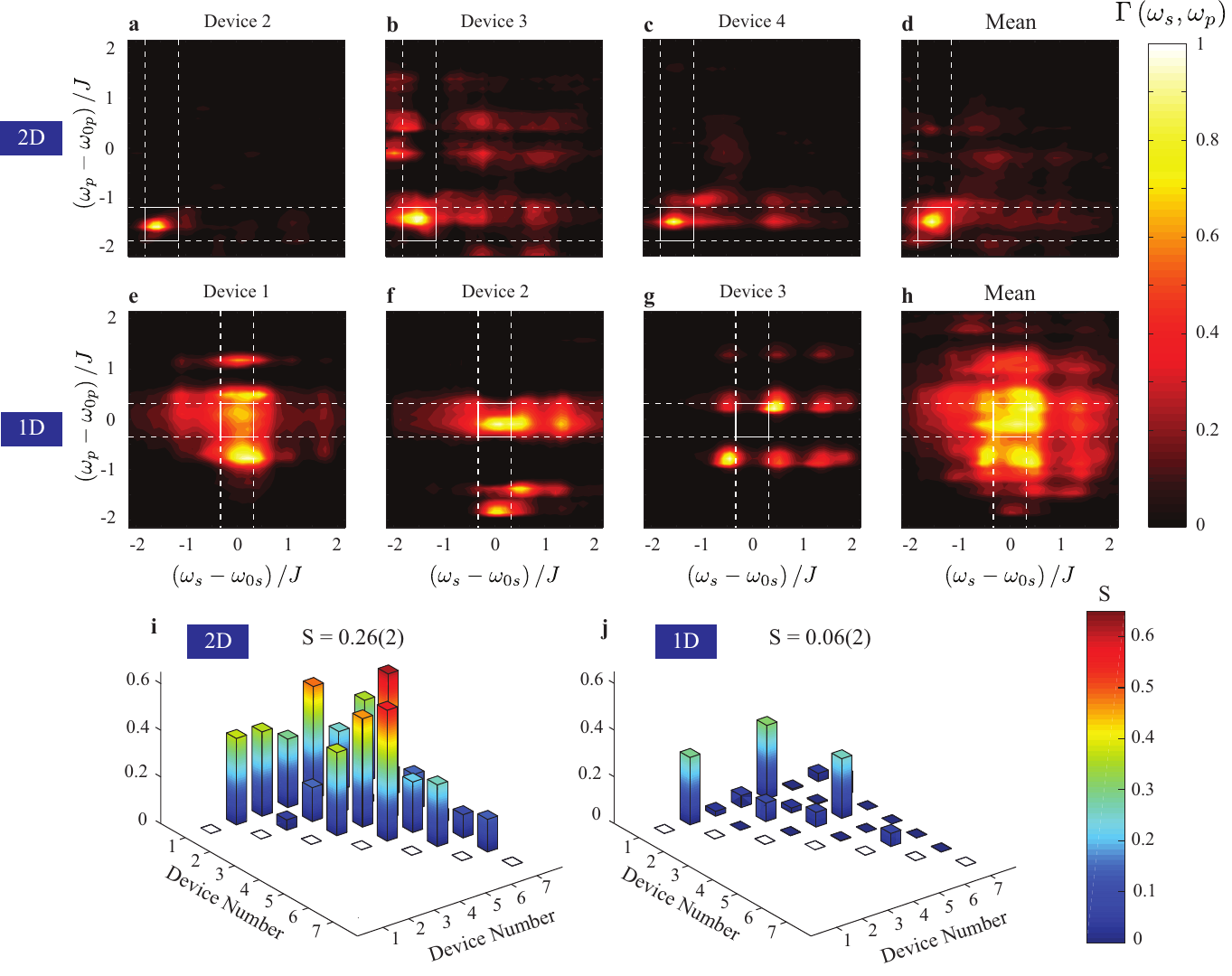}
\caption{
\textbf{Robustness of spectral correlations between pump and signal photons.} \textbf{a-c} Measured $\Gamma\left(\ws,\wp\right)$, for three different 2D topological devices and, \textbf{d} the mean, measured over 7 devices (additional plots in Extended Data Fig. 1). The plots are very similar in the CW edge band region (marked by the solid white box) where the maximum generation of photons occurs. \textbf{e-g} Measured $\Gamma\left(\ws,\wp\right)$, for three different 1D devices and, \textbf{h} the mean, measured over 7 devices (additional plots in Extended Data Fig. 1). There is no region of plot which is similar across all devices. The highlighted region shows the mid-band, $\left|\w_{\mu} -\omega_{0,\mu}\right| < 0.25 J$ where transmission is maximum (see Extended Data Figure 2). \textbf{i,j} Similarity (S) of $\Gamma\left(\ws,\wp\right)$ between the edge band regions of different 2D devices and mid-band regions of 1D devices, respectively.  Because of the topological robustness, edge bands achieve a much higher similarity across devices. The error in similarity measurement for each device pair is less than $3\%$, not shown in the figure.
}
 \end{figure*}

Edge states are topologically protected, quasi-1D waveguides confined to the lattice boundary. Therefore, to benchmark the robustness of these edge channels, we compare them with the topologically trivial 1D waveguides of coupled ring resonators (CROWs, Fig.1f) \cite{Yariv1999, Mittal2014}. The main advantage of CROWs over single ring devices is that they increase the length of SFWM interaction and therefore, the intensity of generated photons, without reducing their bandwidth \cite{Morichetti2011, Davanco2012, Kumar2014}. However, unlike edge states, CROWs are not protected against disorder which can significantly affect the photonic mode structure (see ref. \cite{Mookherjea2008} and Extended Data Fig. E2) and result in device-to-device variations in the spectrum of generated photons. In the following, using measurements over many devices, we show that the topological robustness of our source manifests as a robustness in the spectrum of generated photons and it outperforms the trivial 1D devices.

Figures 4a-c show $\Gamma\left(\ws, \wp \right)$, i.e., the spectrum of generated photons as a function of pump frequency, for three different 2D devices and Fig.4d shows the mean measured over 7 devices (additional data in Extended Data Fig.E1). These devices were designed to be identical but fabrication disorder leads to random variations in the ring resonance frequencies, coupling strengths as well as hopping phases. Nevertheless, as we saw earlier, for all devices the maximum number of photons are always generated in the CW edge band $\left(\ws-\wos \simeq -1.5J \right)$, with pump frequency also in the CW edge band $\left(\wp-\wop \simeq -1.5J \right)$. Therefore, in the CW edge band region (highlighted by dashed white lines), $\Gamma\left(\ws, \wp \right)$ is very similar for all devices. In contrast to edge bands, the spectrum of generated photons in the bulk band differs significantly from one device to the other because it is susceptible to disorder.

Fig.4e-g show similar measurements on 3 different topologically trivial 1D devices and Fig.4h shows the mean measured over 7 devices. As expected, $\Gamma\left(\ws, \wp \right)$ varies markedly from device-to-device, meaning that the spectral correlations between the pump and the generated photons are completely random because of the randomness in the photonic mode structure induced by fabrication disorder. Therefore, given a 1D device, the spectrum of generated photons is not known a priori for any pump frequency. To further quantify and compare the spectral correlations in 2D and 1D devices, we calculate the similarity $S$ between $\Gamma\left(\ws, \wp \right)$ measured on two different devices $(n,n')$, defined as
\begin{equation}
S_{n,n'} = \frac{\left[\int \int d\wp d\ws \sqrt{ \Gamma_{n}
\Gamma_{n'}}\right]^2} {\int \int d\wp d\ws
\Gamma_{n} \int \int d\wp d\ws \Gamma_{n'}}.
\end{equation}
For the 2D devices, we chose the frequency integration interval to cover the CW edge band region $\left[-1.75J, -1.25J \right]$ which is robust against disorder and where the maximum number of photons are generated. For the 1D devices, we choose the mid-band region $\left[-0.25J, +0.25J \right]$ where the pump transmission is maximum (see Extended Data Fig. 2) and for a fair comparison with 2D devices, we choose the same bandwidth of $0.5 J$ as we did for the edge region. These regions of interest are highlighted by white dashed lines in Fig.4a-h. For the 2D system, the average similarity across all devices is $0.26(2)$ whereas for the 1D system it is only $0.06(2)$ (Fig.4i,j). These measurements clearly demonstrate the advantage offered by the topological robustness of our 2D system in engineering the photonic mode structure and consequently, the spectrum of generated photons.

\begin{figure}
\centering
\includegraphics[width=0.49\textwidth]{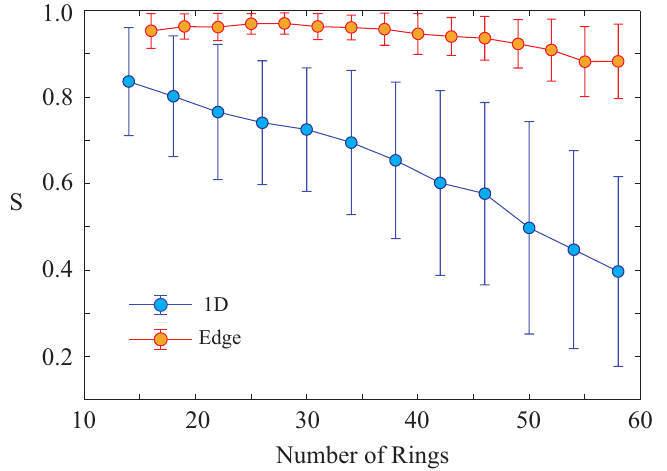}
\caption{
\textbf{Similarity scaling as a function of device size.} Simulated similarity, for moderately disordered ($V = 0.5 J$) 2D and 1D systems, as a function of the number of resonators travelled from input to the output port. Because of topological protection, the 2D system achieves much higher similarity than the trivial 1D system. The results are averaged over 50 realizations of disorder. The error bars represent standard deviation of similarity across different realizations. The solid lines are to guide the eye.
}
\end{figure}

The observed robustness and similarity in our 2D topological devices are remarkable given the fact that our system suffers from a very strong on-site potential disorder, comparable to the edge bandwidth \cite{Mittal2014}. To put our work into perspective and indicate potential future directions regarding scalability, we numerically compare 1D and topological 2D system, for slightly smaller disorder (Fig.5). We observe that the average similarity for the 2D topological system is more than $90\%$ and decreases only marginally as the system size increases. In comparison, for the 1D CROW, the similarity decreases rapidly with system size, approaching $\sim 40\%$ for 60 ring devices. The robustness of topological systems is also evident in the standard deviation of the similarity across different devices, which is significantly smaller compared to that of 1D CROWs. This indicates that with moderate disorder, high visibility two-photon and multi-photon interference \cite{Spring2017} should be possible with photons generated by different 2D topological sources.

In summary, we have demonstrated a topological source of quantum correlated photon pairs where the spectral correlations are robust against fabrication disorder. This is an enabling step towards on-chip, scalable sources of heralded and entangled photons with identical spectra, for applications in quantum information processing and quantum communications. While this demonstration uses devices with relatively high propagation loss $\left(\sim 1 \text{dB/cm} \right)$, recent developments of ultra low-loss photonic platforms $\left(\sim 10^{-3} \text{dB/cm} \right)$ \cite{Bauters2011, Moss2013} can lead to orders of magnitude improvement in the source brightness. Moreover, such low-loss platforms would enable quantum-limited topological amplifiers, where the four-wave mixing gain is required to exceed the propagation losses \cite{Peano2016}.

On a more fundamental level, we have demonstrated a robust route to manipulate the mode structure of the electromagnetic vacuum fluctuations using topological photonics. This can have far reaching implications in engineering light-matter interactions in the quantum regime. We expect intriguing consequences to emerge from application of these ideas to a wide range of physical phenomena, such as, spontaneous emission, super- and sub-radiance, and the Casimir effect.


~~\\
\noindent
\textbf{Acknowledgements:} This research was supported by AFOSR-MURI FA9550-14-1-0267, YIP-ONR, Sloan Foundation and the Physics Frontier Center at the Joint Quantum Institute. We thank Vikram Venkata Orre for help with the experimental setup, Aziz Karasahin for help with the SEM, Tobias Huber and Dirk Englund for fruitful discussions and Q. Quraishi for kindly providing us the nanowire detectors.\\

\noindent
\textbf{Author Contributions:} S.M. and M.H. conceived the idea. S.M. performed the numerical simulations and the experimental measurements. E.G. contributed to source characterization. M.H. supervised the project. All authors contributed to analyzing the data and writing the manuscript.\\

\noindent
\textbf{Data Availability:} The data that support the findings of this study are available from the corresponding author on reasonable request. Correspondence and request for materials should be addressed to S.M. (mittals@umd.edu).


\renewcommand{\theequation}{E\arabic{equation}}
\setcounter{equation}{0}

\newpage
\noindent
\textbf{Methods}\\

\noindent
\textbf{Simulation of Signal/Idler Spectra}\\
In this section, we describe the method used to simulate the spectrum of the generated signal and idler photons in response to a strong pump field. We follow the approach described in \cite{Ong2013, Chen2011}. As described in eq. (1) of the main text, the linear, uncoupled evolution of the pump, signal and idler fields is governed by the Hamiltonian $H_{\text{L}}$, given as
\begin{eqnarray} \label{HL}
H_{\text{L}} &=&  \sum_{m,n} \omega_{0\mu} ~\ad_{m,\mu} \a_{m,\mu} \\
\nonumber    &-& J_{m,n} \left( ~\ad_{m,\mu} \a_{n,\mu} e^{-i \phi_{m,n}} + ~\ad_{n,\mu} \a_{m,\mu} e^{+i \phi_{m,n}} \right).
\end{eqnarray}
Here $\mu = p, s, i$ and corresponds to the pump, signal or idler fields. The nonlinear SFWM process which couples the pump, signal and idler fields is described by the Hamiltonian (eq. (2) of main text)
\begin{equation} \label{HNL}
H_{\text{NL}} = ~\eta  \sum_{m} \left( \ad_{m,s} ~ \ad_{m,i} ~ \a_{m,p} ~\a_{m,p} - \ad_{m,p} ~\ad_{m,p} ~\a_{m,s} ~\a_{m,i} \right).
\end{equation}
Note that this Hamiltonian is local in lattice site index $m$. We assume that the pump field is much stronger than the signal and idler fields and therefore, the evolution of pump field is very well described by the linear Hamiltonian. However, the pump field depletes because of the intrinsic waveguide scattering losses $\left( \kin \right)$, which we include in our simulation. Using the input-output formalism and rotating-wave approximation, we can write the coupled equations describing the steady-state pump field amplitudes, for frequency $\wp$, as
\begin{eqnarray}\label{Evol_Pump}
-i \wp ~\a_{m,p} &=& i \left[H_{L}, \a_{m,p} \right] - \kin ~\a_{m,p} \\
\nonumber        &-& \left( \delta_{m,I} +  \delta_{m,O} \right) \kex ~\a_{m,p} - \delta_{m,I} \sqrt{2 \kex} ~\a_{in,p}.
\end{eqnarray}
Here $\kex$ is the coupling strength of the lattice to input/output waveguides and $\a_{in,p}$ is the input pump field. The input and output waveguides are coupled to the lattice at sites indexed by $I, O$.

Given the pump field amplitudes calculated using \ref{Evol_Pump}, we can write the coupled equations describing the steady-state signal and idler fields amplitudes in the lattice as
\begin{eqnarray}
-i \w_{\mu} ~\a_{m,\mu} &=& i \left[H_{L} + H_{NL}, \a_{m,\mu} \right] - \kin ~\a_{m,\mu} \\
\nonumber               &-& \left( \delta_{m,I} +  \delta_{m,O} \right) \kex ~\a_{m,\mu} - \delta_{m,I} \sqrt{2 \kex} ~\a_{in,\mu},
\end{eqnarray}
where $\mu = s, ~i$. These equations include the nonlinear FWM interaction Hamiltonian of (\ref{HNL}) which couples the signal and idler fields to the pump fields. Also, for a particular choice of frequencies $\wp$ and $\ws$, energy conservation fixes the idler frequency $\wi$.

Using these coupled equations for the pump, signal and idler frequencies, we calculate their field amplitudes in the lattice. Then, the signal/idler fields at the output of the lattice are calculated using the input-output formalism as
\begin{equation}
\a_{out,\mu} = \sqrt{2 \kex} ~\a_{O,\mu},
\end{equation}
where $O$ is the index denoting the lattice output site. We can now define the spectral correlation function (SCF) $\Gamma \left(\ws, \wp \right) = \left|\a_{out,s} \right|^{2}$. This is essentially the spectrum of generated signal photons as a function of pump frequency. Note that because of the energy conservation relation $2\wp = \ws + \wi$, this SCF fully characterizes the spectral correlations of the SFWM process. In other words, using $\Gamma \left(\ws, \wp \right)$, we can easily calculate $\Gamma \left(\ws, \wi \right)$, the joint-spectral density of signal and idler photons.\\

\noindent
\textbf{Experimental Setup}\\
The devices used in this experiment were fabricated using CMOS compatible silicon-on-insulator technology. The ring resonator waveguides are approximately 510 nm in width and 220 nm in height, and at telecom wavelengths ($\sim$ 1550 nm), support only a single mode with transverse electric field. The coupling strength $J$ between the resonators was measured to be 32(1) GHz and the free-spectral range (FSR) was $\approx$ 1035 GHz. The on-site disorder potential $V$ which is a result of the different ring resonance frequencies was estimated to be 27.5 GHz and the disorder on the hopping phase was 0.1. Additional details of the fabrication process and disorder characterization are available in Refs. \cite{Hafezi2013, Mittal2014}.

To generate correlated photons using SFWM process, we pumped the lattice with a telecom band, tunable, CW laser (Santec). The output of the laser was amplified using an erbium-doped fiber amplifier (EDFA) and a tunable band-pass filter was used to cut down the spontaneous emission (ASE) generated during amplification. The pump was coupled to the lattice using grating couplers, with a coupling loss of $\approx$ 5 dB per coupler. At the output of the lattice, tunable band-pass filters were used to remove the pump band, with a rejection exceeding 120 dB. The signal, pump and idler bands were separated by one FSR. To measure the spectrum of generated signal and idler photons, we used two monochromators with a bandwidth of $\approx$ 6 GHz along with two superconducting nanowire single photon detectors (PhotonSpot). The second-order correlation function measurements were done using time-correlated single photon counting system (HydraHarp). \\

\noindent
\textbf{Source Characterization}\\
We use second-order correlation measurements to characterize our source \cite{Eisaman2011, Fortsch2013}. The temporal correlations between signal and idler photons are analyzed using the cross-correlation function $g^{(2)}_{s,i} \left(\tau\right)$ which is given as
\begin{equation}
g^{(2)}_{s,i} \left(\tau\right) = \frac{P_{s,i}\left(\tau\right)}{P_{s} P_{i}}.
\end{equation}
Here $P_{s,i}$ is the probability of detecting a signal photon at time $t$ followed by the detection of an idler photon in the time interval $\left[t+\tau-\frac{\tau_{c}}{2}, t+\tau+\frac{\tau_{c}}{2} \right]$ and $\tau_{c}$ (here 50 ps) is the coincidence time-window. $P_{s}$ and $P_{i}$ are the probabilities of detecting individual signal or idler photons and the product $P_{s} P_{i}$ is the probability of detecting accidental coincidences. We observe that $g^{(2)}_{s,i} \left(\tau\right) \approx 80$ around $\tau = 0$ which implies that the generation of signal and idler photons is strongly correlated. The mean of $g^{(2)}_{s,i} \left(\tau\right)$ around $\tau = 0$ corresponds to actual coincidence counts whereas its mean at $\left| \tau \right| >> 0$ corresponds to accidental counts $\left(P_{s} P_{i}\right)$. Their ratio (CAR) is commonly used as a measure of the signal-to-noise ratio of a source. We measure a maximum CAR $\approx$ 42 when $g^{(2)}_{s,i} $ is averaged over 300 ps (the width of the correlation peak).

The quantum nature of a source can be demonstrated using the second-order auto-correlation function $g^{(2)}_{a} \left(\tau\right)$ which is a measure of antibunching of photons \cite{Eisaman2011, Fortsch2013}. Quantum sources are distinguished by $g^{(2)}_{a} \left(0\right) < 1$ which suggests that the normalized probability of getting two simultaneous photons is low. In the case of correlated photon pairs, the quantum nature is revealed when we measure the conditional auto-correlation function $g^{(2)}_{a,h} \left(\tau\right)$ for signal photons heralded (conditioned) by the detection of idler photons. For this measurement, we use the Hanbury Brown-Twiss (HBT) setup where we place a beam-splitter in the path of signal photons and the outputs of the beam-splitter are connected to two detectors $\left( s_{1} ~\text{and} ~s_{2} \right)$. The idler photons impinge on a third detector ($i$) which heralds the arrival of signal photons. Then the conditional auto-correlation function $g^{(2)}_{a,h} \left(\tau\right)$ for signal photons, conditioned on the detection of idler photons is defined as
\begin{equation}
g^{(2)}_{a,h} \left(\tau\right) = \frac{P_{s_{1},s_{2},i}\left(\tau\right)}{P_{s_{1},i} P_{s_{2},i}}.
\end{equation}
Here $P_{s_{1},s_{2},i}\left(\tau\right)$ is the probability of detecting two heralded signal photons separated by a time $\tau$ and $P_{s_{1,2},i}$ is the probability of detecting individual heralded signal photons. These probabilities are normalized by the probability of idler (heralding) photons. Therefore, $g^{(2)}_{a,h} \left(0\right) = 0 = P_{s_{1},s_{2},i}\left(0\right)$ indicates that the probability of having two pairs of signal and idler photons at the same time is zero. We measure $g^{\left(2 \right)}_{a,h} \left(0 \right) = 0.20 \left(8 \right)$ which is a signature of a good source of heralded single photons.

\newpage

\renewcommand{\thefigure}{E\arabic{figure}}
\setcounter{figure}{0}

\newpage
\linespread{1.0}
\begin{figure*}
 \centering
 \includegraphics[width=0.95\textwidth]{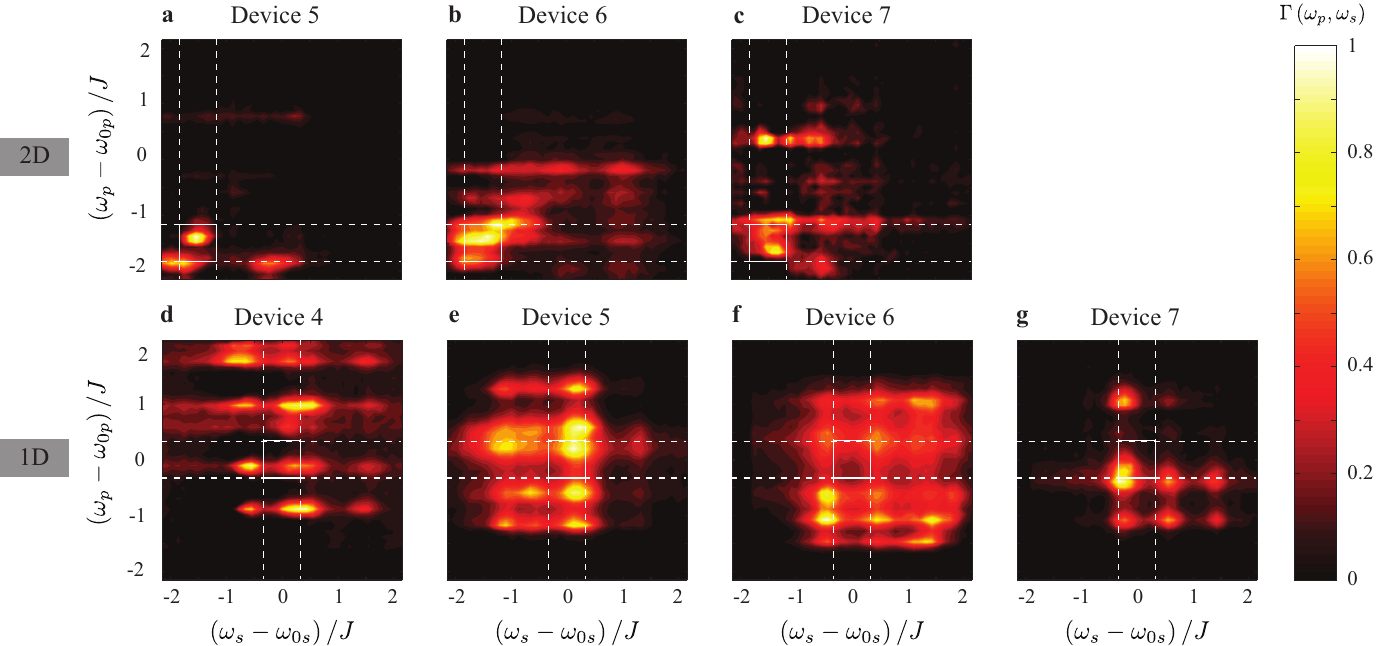}
 \caption{\textbf{Extended Data: Spectral correlations.} \textbf{a-c} Measured $\Gamma\left(\ws,\wp \right)$ on 3 different 2D devices, and \textbf{d-g} on 4 different 1D devices, in addition to those presented in Fig.3 of the main text. CW edge bands for the 2D devices and mid-band for the 1D devices are highlighted. }
 \label{fig:E1}
\end{figure*}
\linespread{2.0}

\newpage
\linespread{1.0}
\begin{figure*}
 \centering
 \includegraphics[width=0.95\textwidth]{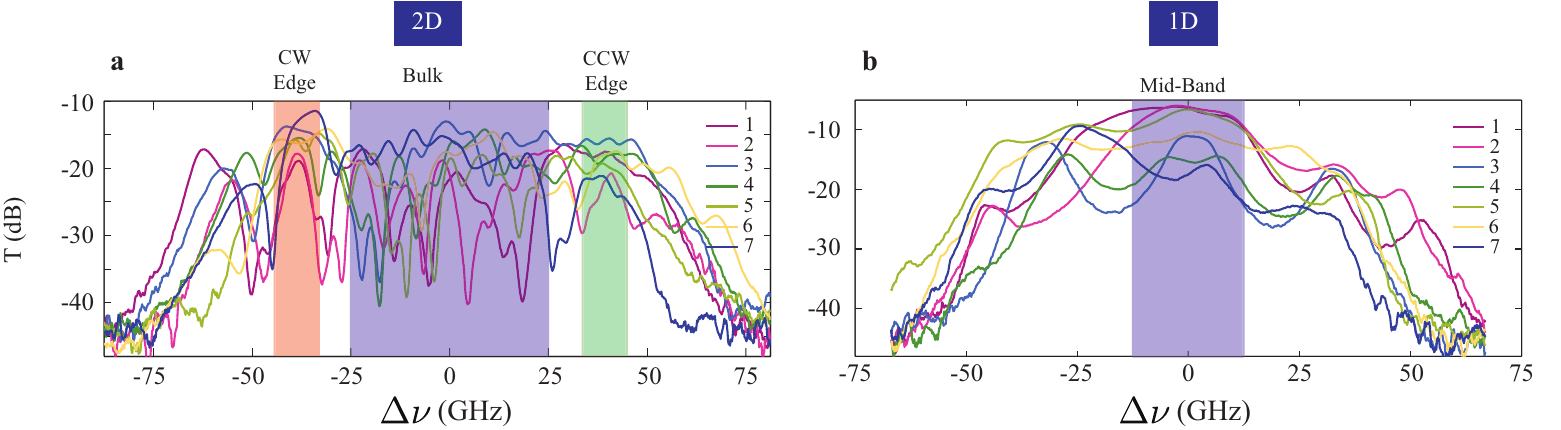}
 \caption{\textbf{Extended Data: Transmission spectra of 2D and 1D devices.} Measured transmission spectra \textbf{a} for different 2D and \textbf{b} for different 1D devices. The shaded regions highlight the edge and the bulk bands for the 2D system and mid-band for the 1D system. For the 2D devices, the CW and the CCW edge bands show reduced variations in the transmission compared to that in the bulk band. These spectra have been shifted along the frequency axis to superpose them, using an algorithm based on transmission and delay measurements, as detailed in Ref.\cite{Mittal2014}.}
 \label{fig:E2}
\end{figure*}
\linespread{2.0}

\newpage
\linespread{1.0}
\begin{figure*}
 \centering
 \includegraphics[width=0.46\textwidth]{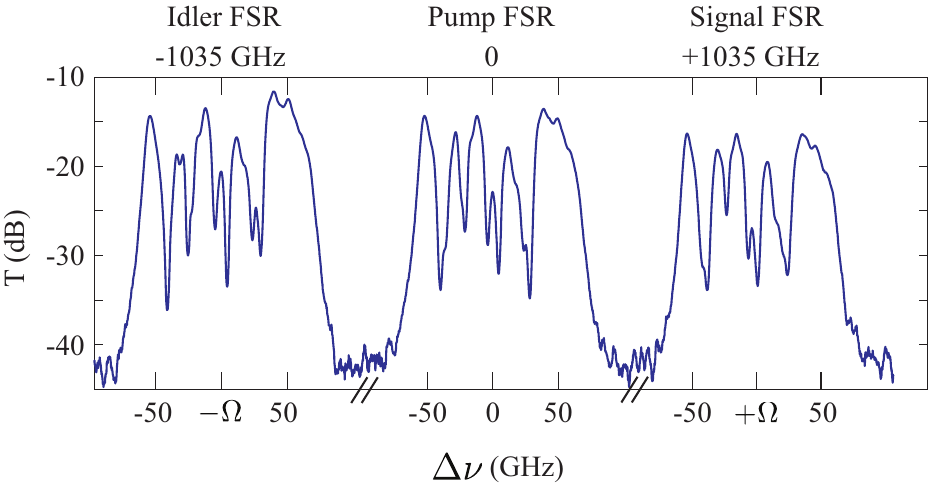}
 \caption{\textbf{Extended Data: Transmission spectrum of a 2D device across three FSRs.} Measured transmission spectrum in the pump, signal and idler FSRs, corresponding to Fig.2 of the main text. As mentioned earlier, the transmission spectrum of the device, i.e., the structure of two edge bands separated by a bulk band, repeats every FSR. Fig.\ref{fig:E3} shows the measured spectrum over the signal, pump and the idler bands, for the 2D device reported in Fig.2 of the main text. The shape of the transmission spectrum in these FSRs is indeed identical. The small variation in the overall transmission across bands is because of the frequency response of the grating couplers. }
 \label{fig:E3}
\end{figure*}
\linespread{2.0}

\newpage
\linespread{1.0}
\begin{figure*}
\centering
\includegraphics[width=0.78\textwidth]{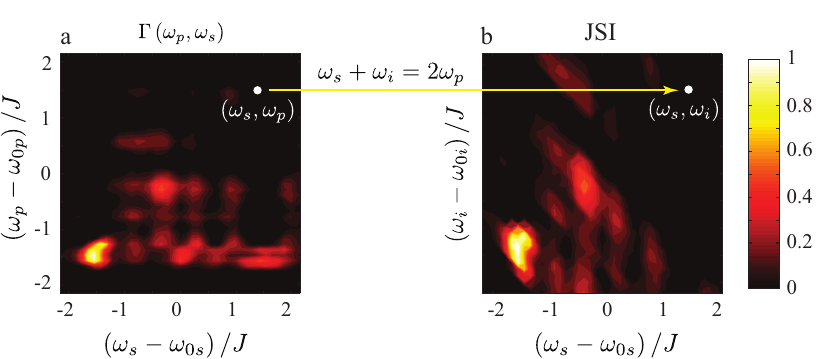}
\caption{
\textbf{Extended Data: Joint-Spectral Intensity.} \textbf{a} Measured $\Gamma\left(\ws,\wp\right)$ (see Fig. 2 of the main text), i.e., the intensity of generated signal photons at frequency $\ws$ as a function of pump frequency, $\wp$. Each point on this plot represents a particular $\ws$ and $\wp$. Using energy conservation, we can calculate the corresponding idler frequency at each point as $\wi = 2\wp - \ws$. Therefore, we can easily rescale the y-axis of the plot and calculate the joint-spectral intensity (JSI, see ref.\cite{Spring2017, Ong2013}) between the signal and idler frequencies, as shown in \textbf{b}. Note that this rescaling works only for a continuous-wave pump because for a pulsed pump source, the above energy conservation relation holds only up to the spectral bandwidth of the pump, signal and idler photons. Also, this measurement inherently assumes that the generated signal and idler photons are correlated. In our experiment, using CAR measurements and direct measurements of the signal and idler spectra (in Fig.2,3 of main text), we verified that the signal and idler photons are indeed correlated. The main advantage of such a spectral correlation measurement between the pump and the signal (or idler) photons is that it is fast and, for a continuous-wave pump, is equivalent to the JSI measurement.
}
\label{fig:E4}
\end{figure*}
\linespread{2.0}

\end{document}